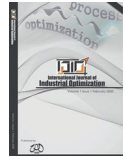

# Advancements in Recommender Systems: A Comprehensive Analysis Based on Data, Algorithms, and Evaluation

Xin Ma [a,1,*], Mingyue Li [b,2], Xuguang Liu [b,3]

[a] Information Resource Management, Business School, Nankai University, China
[b] Software Department, Eco-Chain, Beijing Xiaomi Mobile Software Co., China
[c] Algorithm Development Department, Platform Operation and R&D Center, Jingdong Retail, China
[1] xxin_ma@163.com; [2] li_mingyue@outlook.com; [3] 571773795@qq.com
* Corresponding Author

ARTICLE INFO    ABSTRACT (10PT)

Systematically reviewing and analyzing the current research status of recommender systems (RSs) in the context of emerging technologies, new scenarios, and diverse user needs, is crucial. This helps to accurately grasp the development direction of RSs, strengthen their fundamental research, and strive for their sustainable development. Using 286 research papers collected from Web of Science, ScienceDirect, SpringerLink, arXiv, and Google Scholar databases, a systematic review methodology was adopted to review and summarize the current challenges and potential future developments in data, algorithms, and evaluation aspects of RSs. It was found that RSs involve five major research topics, namely algorithmic improvement, domain applications, user behavior & cognition, data processing & modeling, and social impact & ethics. Collaborative filtering and hybrid recommendation techniques are mainstream. The performance of RSs is jointly limited by four types of eight data issues, two types of twelve algorithmic issues, and two evaluation issues. Notably, data-related issues such as cold start, data sparsity, and data poisoning, algorithmic issues like interest drift, device-cloud collaboration, non-causal driven, and multitask conflicts, along with evaluation issues such as offline data leakage and multi-objective balancing, have prominent impacts. Fusing physiological signals for multimodal modeling, defending against data poisoning through user information behavior, evaluating generative recommendations via social experiments, fine-tuning pre-trained large models to schedule device-cloud resource, enhancing causal inference with deep reinforcement learning, training multi-task models based on probability distributions, using cross-temporal dataset partitioning, and evaluating recommendation objectives across the full lifecycle are feasible solutions to address the aforementioned prominent challenges and unlock the power and value of RSs. The collected literature is mainly based on major international databases, and future research will further expand upon it.

## 1. Introduction

With the widespread adoption of the internet and the rapid growth of digital content, extracting valuable information from massive datasets has become increasingly challenging for consumers [1, 2]. Serving as a crucial complement to search engines, RSs aim to provide personalized recommendations, content, and services based on users' interests and preferences by modeling information behavior and content features from historical data. This helps to deconstruct and predict users' vague information needs, assisting their decision-making [3]. Significant research on RSs has been conducted in both industry and academia. Advanced information and data science technologies have been widely applied to mitigate data sparsity's impact on recommendation quality and user experience [4-8]. Furthermore, application scenarios and programs have greatly expanded [9-12], with the value of recommendations being widely recognized and validated, enhancing their social impact.

New technologies, emerging scenarios, and diverse needs present significant opportunities and potential challenges for the development of RSs. First, new technologies like generative AI can empower RSs by understanding user needs and providing accurate personalized recommendations, but it also raises privacy and ethical concerns, such as data security and algorithmic transparency [1]. For instance, ChatGPT lowers the cost of creating false content, potentially harming social stability. Second, new scenarios like the Internet of Things and Metaverse expand the breadth and depth of RS applications [14], but introduce complex user interactions, data integration and utilization challenges [15], system design and maintenance difficulties [16], and significant security and privacy issues [17]. Third, users' expectations for recommendation services are becoming more diverse. They demand not only accuracy but also broader social responsibility and sustainable development, including fair information filtering and ranking [13, 18], and the avoidance of filter bubbles [19].

In summary, it is essential to explore and assess the current state of RSs research in light of new technologies, emerging scenarios, and diverse needs to accurately grasp the direction of RSs research and provide guidance for its sustainable development. Although existing review articles have organized RSs research around various interests, such as online courses [20], privacy protection [21], and cancer management [10], there are still some gaps: (1) Most papers focus on specific topics within a single domain, such as interpretable information [22], blockchain [4], or healthcare [14], lacking a comprehensive evaluation of RSs, especially the prominent challenges in data, algorithms, and evaluation amid new technologies, scenarios, and diverse needs. (2) Many reviews are narrative or commentary-based, which, despite organizing extensive literature, are relatively subjective and lack clear methodologies and quantitative analysis, making data and results difficult to reproduce. While some studies use systematic review methods, they mostly focus on technical classifications of RSs [3, 14, 23, 24].

To comprehensively present the latest trends in RS research, provide scientific insights for scholars in information science, management science, and computer science, as well as practical guidance for industry practitioners innovating and applying RSs, this study uses a systematic review method to review and organize 286 RS research papers to address the research question: what severe challenges do RSs face in terms of data, algorithms, and evaluation in the context of new technologies, emerging scenarios, and diverse needs?

## 2. Review Principles and Methods

### 2.1. Review Principles

This study established three principles to ensure an objective paper review process: First, two experienced RS researchers independently handled paper screening, data processing, and coding. A third evaluator resolved any disagreements. Second, all evaluators received standardized training to ensure a consistent understanding of screening criteria and coding rules. Third, any conflicts of interest between evaluators and the papers under review were promptly disclosed, and the first author of the study decided whether to replace the evaluator if necessary.

### 2.2. Systematic Review Method

This study employed a systematic review method to comprehensively organize the literature on RS research. Systematic review is an approach that systematically evaluates literature using a predefined and explicit process [25, 26]. It focuses on specific questions and follows mature procedures, such as literature inclusion and exclusion criteria, quality assessment, evidence analysis & synthesis, and results reporting & summarization (PRISMA guidelines [26]). Unlike narrative and commentary reviews, this approach is systematic, transparent, and replicable [27], ensuring scientific rigor and reliability in knowledge innovation. As shown in Fig.1, this study followed the PRISMA guideline's four-stage process for collecting and screening relevant literature.

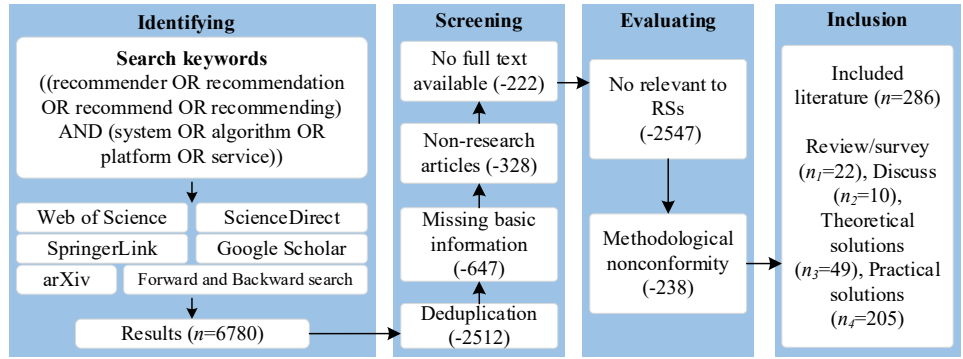

**Fig. 1.** The literature inclusion process for RSs following the PRISMA guidelines

### 2.2.1. Literature Retrieval

This study conducted a three-step retrieval of RS literature. First, searches were performed in multidisciplinary databases including Web of Science, ScienceDirect, and SpringerLink using predefined terms from the past five years: ([recommender, recommendation, recommend, recommending]) AND ([system, algorithm, platform, service]). Second, gray literature was supplemented by searching arXiv and Google Scholar. Finally, based on the review papers obtained from the first two steps, forward and backward searches [28] were conducted to further collect literature. The search terms were initially constructed based on relevant research reports and literature, and finalized after review by three experts in the RS field.

### 2.2.2. Literature Screening and Evaluation

The search, conducted by the end of October 2023, yielded 6780 papers. Using the Web of Science Core Collection as a baseline, papers were screened according to the criteria in Table 1. Exclusions were made for duplication (-2512), missing information (-647 papers), non-research content (328 papers), lack of full text (222 papers), irrelevance to RS (2547 papers), and methodological irregularities (238 papers). Ultimately, 286 papers were selected for further analysis, aligning with recommendations for systematic reviews [25, 29].

**Table 1.** Literature Exclusion Criteria and Descriptions for RS Research

| Criteria | Description |
| --- | --- |
| Duplication | 1. Identical papers across different databases. 2. Papers with the same authors, topics, and high content overlap, such as conference papers and journal articles. |
| Missing basic information | Incomplete information such as names, affiliations, abstracts, and journal sources. |
| Lacking full text | Unable to access the full text of the paper. |
| Non-research articles | Papers not related to scientific research. For example: conference reviews, call for papers' announcements, special issue introductions, etc. |
| Unrelated to RSs | 1. The keyword "RSs" appears multiple times in the text but is not directly related to RSs. 2. The paper does not focus on RSs in terms of review, survey, discussion, or problem-solving (theoretical or practical): 1) RSs are only used as an example in the text; 2) RSs are mentioned only in relation to future research, perspectives, or needs in the text; 3) RSs are cited only once in the text; 4) RSs appear only in the title, abstract, keywords, or references. |
| Methodologically non-standard | The research design, experimental setup, data collection, and analysis process of the paper are unreliable or invalid. For example: not specifying the parameters of the |

control algorithm, not open-sourcing data or code, having too few references, or using outdated control algorithms.

### 2.2.3. Data Coding

Key categories extracted and coded from the collected papers include: (1) basic data such as affiliation, application domains [3], and retrieval time; (2) research types (review/survey, discussion, theoretical solutions, practical solutions) [28]; (3) recommender techniques [30]; (4) primary data issues explored; (5) primary algorithmic issues explored; and (6) primary evaluation issues explored. Given the lack of widely accepted coding rules for RS data, algorithms, and evaluations, an inductive approach was used for coding.

## 3. Bibliometric Analysis

### 3.1. Distribution of Literature Types

Among the 286 collected papers, 215 are journal papers and 71 are conference papers. The journal articles appeared in 132 different English-language journals, with IEEE Access (28 papers) and Applied Sciences Basel (13 papers) being the most prolific. The top three active fields in RSs are computer science (83%), engineering (31%), and mathematics (27%). In conferences, papers were sourced from 13 different events, with RecSys having the most publications (19 papers), followed by SIGIR and CIKM (15 papers each).

### 3.2. Distribution of Geographic Region

Among the 286 collected papers, over 82% are affiliated with institutions in Asia, followed by Europe (189 papers), North America (55 papers), Oceania (21 papers), Africa (21 papers), and South America (8 papers). Analyzing the top two countries by publication volume in each region, China shows significant interest in RSs (121 papers). In contrast, the Americas, Africa, and other Oceania countries show less focus on RSs compared to the United States (46 papers), Italy (31 papers), the United Kingdom (28 papers), and Australia (20 papers).

### 3.3. Distribution of Application Domain

69 papers on applied topics cover six domains: entertainment, health, tourism, internet/e-commerce, education, and social media/other. The most popular areas are entertainment (19 papers), social media/other (14 papers), and internet/e-commerce (13 papers), focusing on enhancing user experience, promoting content dissemination, and boosting sales. Health (11 papers) and education (8 papers) follow, with trends in proactive health [31] and online learning [20]. Tourism (4 papers) garners the least attention, but has potential opportunities in the post-pandemic era due to the increasing demand for personalized cultural and tourism experiences.

### 3.4. Distribution of Hot Topics

Using LDA topic clustering on abstracts from 286 collected papers, we identified five hot research topics: (1) Algorithm Improvements (101 papers), optimizing recommendation algorithms with techniques like deep learning [23, 32], knowledge distillation [33], and reinforcement learning [34] to address data sparsity and scalability issues. (2) Domain Applications (69 papers), designing and implementing RSs in various fields such as health [35], tourism [36], and pest detection [37] to provide targeted services; (3) User Behavior/Cognition (40 papers), studying user behavior patterns, cognitive processes, and subjective experiences to enhance RSs usability and user experience [38, 39]; (4) Data Processing/Modeling (57 papers), boosting data quality and data understanding through the collection, cleaning, and storage of large-scale behavioral data, item attributes, and contextual information [40-42]. (5) Social Impact/Ethics (19 papers), addressing ethical challenges like filter bubbles [19] and privacy breaches [43], and proposing solutions for the trustworthy development of RS technologies.

### 3.5. Distribution of Core Technologies

Among 202 papers on practical solution-oriented topics (excluding empirical studies), we identified 139 types of RS technologies across five categories (Fig.2): collaborative filtering (66 papers), content-based recommendation (34 papers), knowledge-based recommendation (20 papers), hybrid recommendation (61 papers), and context-aware recommendation (21 papers). Collaborative filtering, the most studied, predicts user-item interactions based on existing data and includes model-based and memory-based approaches. Technologies like Bayesian networks [44], clustering [45], reinforcement learning [34, 46], decision trees [47], attention [48], distribution calibration [49], and rating reliability [50] are commonly used to enhance zero-shot or few-shot recommendations. Hybrid recommendation strategies, integrating various technologies, also garnered interest. These include feature-enhanced hybrid, cascade hybrid, embedding hybrid, and mixed hybrid strategies [51-55], using multilayer perceptron [56], factorization networks [57], ChatGPT with prompt engineering [58, 59], meta-controllers [60], and Bayesian ranking [61]. Content-based recommendations leverage multi-source and multi-modal item data to match user interests, employing BiLSTM [62], Transformer [63], and deep trust networks [64] for key attribute extraction.

Research on context-aware and knowledge-based recommendations is limited. Context-aware recommendation integrates contextual information like time and weather using methods such as contextual filtering and context modeling, with technologies like deep autoencoders [31, 65], deep neural networks [66], Zipf's law [67], and edge computing [68]. Knowledge-based recommendation, the least explored area, utilizes domain knowledge, user profiles, and other external information to enhance recommendation specialization, using techniques like graph neural networks [69-70], knowledge graphs [71-73], and semantic web rules [74].

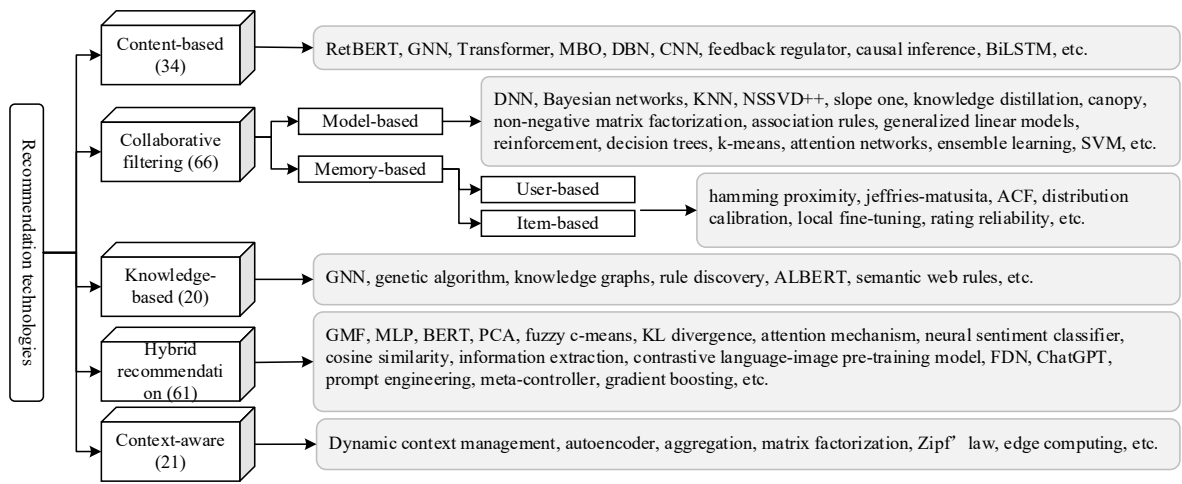

**Fig. 2.** The Classification of Key Technologies in RS Research

## 4. Challenges in Data, Algorithms, and Evaluation of RS Research

### 4.1. Data Problems for RSs

Data forms the foundation of RSs, determining the upper limits of service quality [75]. Personalization in RSs hinges on the reliability, diversity, and scale of user historical behavior, preferences, and interest data. Insufficient or low-quality data can introduce model bias, limiting the effectiveness of recommendations. As identified in the review (Table 2), 87 papers discuss RSs data issues across 8 types grouped into 4 classes. Due to space limitations, this study analyzes three data challenges—cold start, data sparsity, and data poisoning—that may evolve in manifestation, mitigation strategies, and impact scope amid new technologies, scenarios, and diverse demand.

#### 4.1.1. Cold Start and Data Sparsity

Accurate recommendation in RSs relies on sufficient user-item interactions; limited or absent interactions severely impacts recommendation outcomes and diminish user experience [76]. Cold start and data sparsity are critical concerns. Cold start occurs when RSs lack initial or new user and item data, hindering the inference of user preferences [67]. Data sparsity arises from uneven user behavior data, limiting comprehensive modeling and prediction for all users or items. Current research emphasizes integrating multimodal information to enrich historical interactions [77]. Some studies apply deep learning to fuse modalities such as images, text, and audio [63, 78], enhancing RSs' ability to understand and predict user interests. Others incorporate social network data to explore user behaviors and preferences [56], enriching user feature representations. Additionally, research utilizing sensor data and environmental information [65, 68] constructs contextual cues for user behavior, improving RSs' applicability and utility. Despite progress in mitigating cold start and data sparsity, several challenges remain notable. First, traditional modal data like text and images are static, meaning their content or attributes do not change significantly within a period, which hampers meeting users' real-time and interactive needs. Second, advancements in new technologies and scenarios such as brain-computer interfaces and mixed reality [28] are changing how users acquire information and interact [79]. Capturing user behaviors and preferences in virtual environments using physiological signals like eye movements and EEG remains a topic requiring exploration.

Table 2. Literature Exclusion Criteria and Explanations for RS Research

| Primary classification | Secondary classification | Description | Number of papers |
|---|---|---|---|
| Data bias | System bias | Pre-existing interaction design limits data representation of user preferences. | 13 |
| | Misleading bias | Obvious item attributes misrepresent the data's representation of user preferences. | 9 |
| | Feedback bias | Effects such as long tail, herd and exposure lead to imbalanced user feedback data. | 10 |
| Data missing | Data sparsity | The 2D interaction space between the user and the item is missing most of the interaction data. | 22 |
| | Cold start | Data for new scenarios, new users, and new projects is completely missing. | 17 |
| Data noise | Artificial noise | Data noise introduced by subjective user error or historical limitations. | 4 |
| | Data poisoning | Data noise deliberately introduced into the system (e.g. fake reviews of goods). | 11 |
| Data abuse | Data kill | It refers to the recommendation of the same product or service to users under different data transaction conditions. | 8 |

### 4.1.2. Data Poisoning

RSs are essential for web service platforms to manage information overload by modeling complex user-item interactions. However, the openness of these platforms allows any entity to register and leave interaction traces, increasing the risk of data poisoning. Data poisoning involves providing false information or manipulating behaviors, leading to biased or misleading recommendations [80]. This can occur through user poisoning, which manipulates profiles, preferences, and interactions [81], or item poisoning, which alters item content to distort the original evaluation and ranking of items [1, 82]. These behaviors seriously threaten the fairness, accuracy, and credibility of RSs, impacting user experience and platform sustainability. While research has proposed countermeasures such as enhancing data quality monitoring [83], designing robust algorithms [41], strengthening privacy protection [43], and creating transparent RSs [22], challenges remain. First, new technologies (ChatGPT-like models [84]) enhance the adaptability of data poisoning, reducing the effectiveness of existing defenses and leading to a continuous cycle of attacks and defenses. Second, RSs face complex and multi-source data poisoning in cross-domain scenarios like the IoT and metaverse. Addressing user behavior and environmental contexts on complex devices, and enhancing real-time and transferable defenses against data poisoning, require further exploration.

**4.2. Algorithmic Problems for RSs**

Algorithms form the core of RSs, crucially impacting service performance [85]. As data scales and application scenarios expand, the selection, design, and optimization of algorithms become increasingly important. Studying potential challenges and developments in new contexts is vital for accurately understanding user interests, enhancing recommendation quality, and improving user experience. According to the review (Table 3), 142 of the 286 collected papers fully or partially focus on RS algorithms, addressing twelve types of issues across two categories. Similar to Section 4.1, this study specifically analyzes four algorithmic challenges: interest drift, device-cloud collaboration, non-causal drivers, and multitask conflicts.

Table 3. Algorithmic Problems and Their Classification in RS Research

| Primary classification | Secondary classification | Description | Number of papers |
|---|---|---|---|
| Low performance | Cross-domain adaptive | Users' interests migrate between different domains. | 11 |
| | Device-cloud collaboration | Efficient recommendation with limited computing, storage, and bandwidth resources. | 9 |
| | Multitask conflicts | A balance of possible conflicts between users' different interests. | 12 |
| | Low accuracy | Lack of accuracy in recommendations (e.g. low accuracy and recall rates). | 28 |
| | Environmental awareness | Consider the user's environmental information (such as geographic location and device type). | 14 |
| | Interest drift | Users' interests, behaviors and needs change over time, showing temporal and drift. | 8 |
| Low sustainability | Non-causal drivers | Generating recommendations based on non-causal relationships (association rules and user behavior patterns). | 12 |
| | Uninterpretable | Recommendations are difficult to interpret. | 13 |
| | Algorithmic black box | Algorithms obscure the data processing and decision-making process, or users cannot understand the algorithm logic. | 7 |
| | Algorithmic discrimination | Excessive preference or discrimination towards certain users/items (e.g., gender discrimination). | 10 |
| | Information cocoon | Reinforcing existing user interests, reducing exposure to diverse content. | 11 |
| | Privacy security | User privacy data is compromised due to algorithm vulnerabilities or hacker attacks. | 18 |

**4.2.1. Interest Drift**

The mainstream recommendation paradigm ranks and recommends existing items based on user feedback and contextual information [58]. However, due to flaws in model design and data distribution, most RSs designed under this traditional paradigm often underperform in practical deployments. The main reasons include: firstly, traditional RSs often use fixed rules or pattern matching to recommend items, limiting their ability to generate diverse new content in real-time and meet diverse user needs [58]. Secondly, they primarily use implicit feedback [86], which is sparse and fails to clearly express user needs, especially negative ones, limiting system stability and generalization in zero-shot or few-shot tasks. Finally, considering system design and maintenance, traditional RSs often present recommendations directly without sufficient user interaction and feedback [73, 87], making it difficult to perceive changes in user preferences and actively correct biases. While generative recommendation offers a promising new paradigm to address these challenges, focusing on interpretability [59], recommendation effectiveness [88], content generation quality, and diversity [58], practical applications still face numerous challenges. Firstly, existing research often focuses on specific domains or platforms, lacking consideration for diverse experimental contexts. The application scenarios and scope of generative recommendation are not fully explored, and how to apply it effectively across different domains and user groups remains unsolved. Secondly, existing research emphasizes the quality and diversity of generated content, while often neglecting other important metrics such as satisfaction, cost, and social impact. Designing, evaluating, and optimizing generative recommendations while meeting these metrics requires further in-depth research.

**4.2.2. Device-Cloud Collaboration**

With the rise of mobile computing and IoT applications, especially improvements in mobile device performance, computing paradigms are decentralizing [89, 90]. Scholars in RSs focus on coordinating small device-side models and large cloud-side models in three areas: device-side inference (using device-side computing resources for intelligent inference and decision-making [68]), device-side learning (fine-tuning cloud static models with real-time click sequence segments from end-users [49], or aggregating local parameter updates into a cloud model to address limited device resources and privacy concerns), and device-cloud scheduling (optimizing resource utilization and recommendation efficiency by coordinating computing tasks between cloud and device [42, 60]). Device-cloud scheduling, which fully utilizes cloud computing resources for efficient and flexible model computations and inference while relieving device pressure and enhancing RS performance, has become a key focus in current research. However, it still faces challenges. Firstly, as edge devices and computing tasks grow, traditional scheduling struggles with real-time efficiency. There is a need for more efficient mechanisms to cope with the growing computational demands, and adaptive machine learning and pre-trained large language models offer potential solutions. Secondly, emerging scenarios like smart homes and new energy networked vehicles have made user needs more diverse, demanding flexible and intelligent scheduling for personalized resource allocation.

**4.2.3. Non-Causal Drivers**

Current RS research relies on modeling correlations using user historical behavior or user/item features to link high-value content with user needs. However, the real world is driven by causal relationships, and correlations do not imply causation [91]. Relying solely on correlations can compromise fairness, interpretability [22], and generalization [92] of recommendations, limiting the scope of information retrieval. To address this, researchers use causal inference to model and infer causal relationships in RSs. Some studies explore adjusting causal relationships through front-door or back-door methods, using causal graphs and structural functions to identify and reduce recommendation biases [93]. Others employ counterfactual learning, representation learning, and causal discovery to capture the impact of various factors on recommendations, aiming to enhance RSs' interpretability and generalization [94]. Despite the potential of causal inference to improve recommendation effectiveness and utility, several challenges remain. Firstly, user decisions involve numerous factors, making the manual design of causal graphs limited and less applicable for causal discovery [92]. Research is needed on leveraging new technologies to automatically generate and optimize causal graphs from large datasets, accurately modeling complex causal relationships. Secondly, most causal inference models address specific recommendation problems. With diverse user demands and new scenarios, enhancing the universality and flexibility of causal inference is essential to evolving the recommendation environment.

**4.2.4. Multitask Conflicts**

Given the diverse user demands in practical applications, RSs need to handle multiple recommendation tasks simultaneously. Enhancing RS performance in a multitasking environment has become increasingly popular [46]. Multitask learning enables recommendation models to learn multiple tasks simultaneously within a unified framework, such as ranking and click-through rate estimation, improving overall task effectiveness by sharing underlying representations [95]. Significant progress has been made in multitask learning for RSs. For instance, researchers have proposed effective models, including those based on knowledge graphs [57], feature decomposition networks [72], and machine learning approaches [34]. Additionally, there is ongoing exploration of designing multitask learning frameworks to model task correlations [96] and mechanisms for parameter sharing. Despite these advances, several challenges remain. First, the diversity and complexity of tasks can lead to imbalances in model performance. Research is needed on leveraging new technologies to design effective model structures and training strategies to balance the mutual influence and interference between tasks. Second, as the number and variety of tasks increase, ensuring the model's generalization to new tasks and its scalability are major challenges in multitask learning for RSs.

### 4.3. Evaluation Problems for RSs

Evaluation is crucial for RSs, determining the effectiveness of recommendation services [97]. A comprehensive evaluation system assesses performance effectively, guiding algorithm optimization, and reflects real user needs, and enhancing user stickiness. Of the 286 collected papers, only 15 fully or partially focus on RS evaluation, addressing two main issues: offline data leakage (4 papers) and multi-objective balancing (11 papers).

#### 4.3.1. Offline Data Leakage

Due to restricted access to online platform visits, RS research typically relies on pre-collected datasets for offline experimental evaluation [98]. The partitioning of training and test sets is crucial and is commonly done through three methods: random partitioning (randomly selecting test instances from the dataset based on a specified ratio), leave-one-out partitioning (using each user's last interaction as a test instance), and time-based partitioning (selecting test instances based on a specific time point in the series) [99-100]. However, these methods face the issue of offline data leakage, which involves using future information not available during offline training [99, 101-103]. For instance, in leave-one-out partitioning, using the nth interaction of each user for testing may inadvertently include future interactions from user v in the preceding (n-1) interactions of user u. While some studies attempt to build time-aware RSs using timestamps of training instances [104], challenges remain. First, emerging technologies like pre-trained large language models or generative AI empower RSs to capture user interests accurately but also demand larger data scales, complicating time-aware modeling and increasing data leakage risks. Second, existing time-aware models often focus on local temporal aspects for specific users or items but lack globally observable temporal features for comprehensive evaluation, resulting in inevitable data leakage.

#### 4.3.2. Multi-Objective Balancing

Traditional RS research has focused on providing solutions for quickly accessing relevant information, treating recommendations as supervised machine learning problems to accurately predict user preferences and item relevance [104]. Metrics such as precision, mean absolute error, and normalized discounted cumulative gain have been widely used to minimize prediction errors or maximize accuracy. However, focusing solely on accuracy is insufficient; the value of recommendations is equally important [105, 106]. For example, a music RS that only recommends similar music based on a user's listening history may have minimal prediction error but lack engaging recommendations. This realization has shifted the focus to 'beyond accuracy' metrics. Studies are now examining the limitations of traditional accuracy metrics and exploring new ones such as diversity, novelty, and content safety [1, 107]. Other research aims to balance accuracy with these 'beyond accuracy' metrics, proposing methods to optimize comprehensive performance [108, 109]. Despite these advancements, challenges remain with new technologies, scenarios, or diverse user demands. First, as new algorithms and scenarios emerge, traditional metrics struggle to effectively assess their performance, particularly regarding ethical considerations in generative recommendations, requiring adjustments and new metrics. Second, RSs face conflicting goals across stages, such as precision versus diversity [105]. A comprehensive evaluation of RS performance throughout their lifecycle is an area needing further exploration.

## 5. Future Research Implications

New technologies, emerging scenarios, and diverse user demands present significant opportunities for RSs but also challenge traditional RS data, algorithms, and evaluation methods.

Researchers urgently need to examine the shortcomings and potential extensions of existing RS research from multiple perspectives to better adapt to these new trends.

## 5.1. Topics and Technological Perspectives

This study, through a comprehensive survey of RS research, reveals key trends in research topics and recommendation technologies: (1) RSs research covers five major topics: algorithm improvements, domain applications, data processing & modeling, user behavior & cognition, and social impact & ethics. Algorithm improvement is the most prevalent, followed by domain applications, data processing & modeling, and user behavior & cognition, with the least focus on social impact and ethics; (2) collaborative filtering, hybrid recommendation, content-based recommendation, context-aware recommendation, and knowledge-based recommendation constitute a diverse technological framework of 139 types within RSs. These findings indicate that the RS field is flourishing, with diverse research topics and a varied technological landscape offering rich potential for future growth. While deepening research in algorithm improvements, domain applications, data processing & modeling, and user behavior & cognition, it is crucial to enhance focus on social impact and ethics to ensure alignment with ethical standards and societal values. This promotes the comprehensive and sustainable development of personalized recommendations. With emerging technologies like pre-trained large language models, augmented reality, and edge computing, researchers should focus on collaborative RS technologies. This would facilitate cross-technological integration and innovative applications, address diverse user demands and dynamic external environments, and establish intelligent, sustainable, and responsible RSs.

## 5.2. Data Perspective

### 5.2.1. Fusion: Introducing physiological signals to enhance modeling integrity

Adequate data forms the foundation for RSs to understand user preferences, behaviors, and trends, driving personalized recommendation services. This study found significant attention given to enhancing system historical interactions using multimodal data. Traditional modalities can alleviate cold start and data sparsity but lack the dynamism to capture real-time changes in user behavior and preferences. Integrating physiological signals with traditional modal data could address these issues, improving the real-time responsiveness, interactivity, and intelligence of RSs. With the increasing prevalence of wearable devices and smart sensors, like Apple's Vision Pro and Neuralink's brain-machine interfaces, collecting physiological signals is becoming more convenient and real-time. This technical support enables highly interactive and personalized recommendations. Physiological signals provide immediate feedback, reflecting not only users' surface preferences but also their deeper emotional states, cognitive characteristics, and focus of attention [110]. By integrating these signals with traditional modal data, RSs can comprehensively understand user behaviors and accurately capture real-time changes in user needs and interests.

For instance, using EEG data (Fig.3), real-time collection, processing, and analysis of electroencephalography signals from wearable devices or brain-machine interfaces can be integrated with traditional data such as text, images, and video. This enables multidimensional and in-depth modeling and recommendation of user behaviors and preferences.

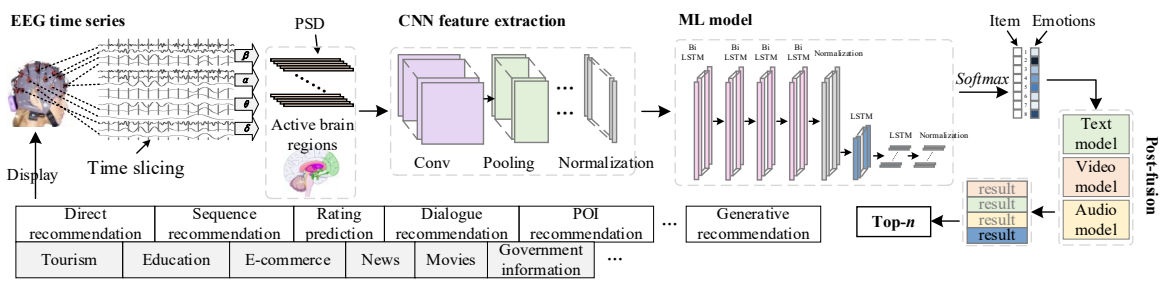

**Fig. 3.** A Research Framework for a Multimodel RS Incorporating EEG Data

**5.2.2. Defense: Modeling user information behavior to defend against data poisoning**

The authenticity and compliance of data are crucial for protecting user interests, complying with legal regulations, and upholding ethical standards, impacting the security and longevity of RSs. This study identifies user poisoning and item poisoning as key data threats, primarily defended against with heuristic-robust methods. However, rapidly evolving technologies and scenarios shorten these methods' lifecycles, intensifying the ongoing "cat-and-mouse" game of attack and defense. Using information retrieval models to deconstruct user behavior offers a promising defense against data poisoning. Compared to complex and varied data poisoning techniques, individual user behaviors are relatively independent and objective [111]. These models, grounded in strong theoretical and practical foundations, can understand user behaviors' motivations and mechanisms, aiding in defense strategy formulation. They can also be continuously optimized with large-scale data training to detect and counter new data poisoning techniques promptly. Moreover, their flexibility and scalability allow integration with other security technologies across different domains and scenarios to build multi-level, multi-dimensional defense systems.

Taking the Berrypicking behavior model [112] as an example (Fig.4), deep learning can model users' dynamic behavior in information retrieval logs, which involves multiple selections, discards, and reselections. This approach can help design intelligent data poisoning detection systems that identify anomalous behaviors and implement defenses, such as filtering out malicious items during interactions.

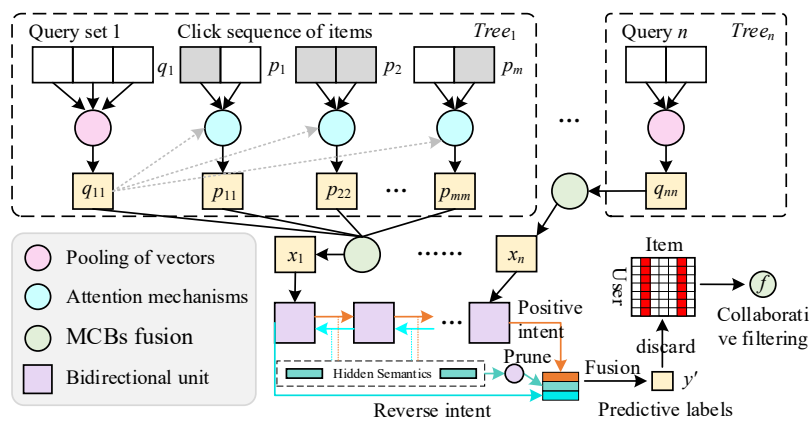

**Fig. 4.** A Research Framework for a RS based on the Berrypicking Model to Prevent Data Poisoning

## 5.3. Algorithmic Perspective

**5.3.1. Validation: Utilizing social experiments to evaluate generative recommendations**

Perceiving and satisfying dynamic user interests and preferences is crucial for RSs to enhance user engagement, build loyalty, and increase revenue. Generative recommendations offer stronger personalized customization, better capturing the evolution of user interests and predicting future behaviors. However, they often overlook diverse contexts, and their application scenarios, scope, practical value, and development directions remain underexplored. Using social experiments to assess generative recommendations in real-world scenarios is a critical research direction. By designing specific scenarios and conditions, social experiments can explore the applicability and effectiveness of generative recommendations across different contexts, comparing performance across cultural backgrounds, user groups, and product types. Additionally, social experiments allow observation and analysis of user behaviors and feedback, helping to understand the impact of generative recommendations on user decisions. They also reveal user acceptance, satisfaction, and concerns about privacy and data security, providing insights for the design and improvement of RSs.

Using controlled field experiments as an example (Fig.5), simulated real-world applications of generative recommendations, such as in psychological counseling and chatbots, can compare experimental and control group data. This analysis assesses the applicability and performance differences between generative and traditional recommendations across various contexts, revealing their strengths and limitations. This approach provides a scientific basis for designing and improving generative recommendations.

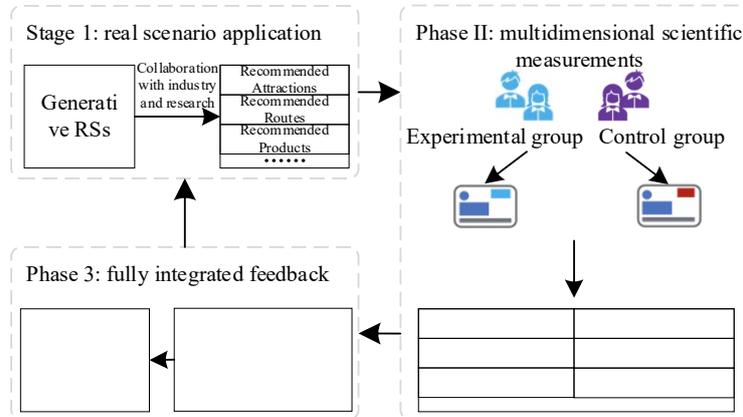

**Fig. 5.** A Research Framework for Evaluating Generative RSs' Effectiveness Using Controlled Field Experiments

### 5.3.2. Adjustment: Fine-tuning pre-trained large models for scheduling device-cloud resources

Combining global learning of cloud-based large models with local optimization of edge-based small models is crucial for coordinating device-cloud resources in RSs. This strategy enhances recommendation efficiency, reduces data transmission and computational costs, and balances personalized recommendations with overall system development. This study identifies three device-cloud coordination strategies: device-side inference, device-side learning, and device-cloud scheduling, with the letter being predominant. However, current scheduling strategies lack the flexibility and intelligence for complex scenarios, tasks, and diverse user needs. Designing lightweight universal schedulers for pre-trained large models is a promising research direction. These schedulers, trained on massive cloud data and high-performance GPU clusters, excel in logical relationships, domain transfer, and cognitive reasoning, making them suitable for real-time recommendation tasks by reducing inference latency and resource consumption. Additionally, lightweight schedulers, typically tens of megabytes in size, impose minimal pressure on edge devices, allowing dynamic adjustment of model parameters based on local user preferences, thus balancing personalized recommendations with system performance.

Taking large language models like ChatGPT as an example (Fig.6), distributed inference can fine-tune model parameters (cloud-side) [113] or distill network structures (device-side), adapting to specific computing resources and user demands. This efficiently schedules device-cloud resources. Interactive learning techniques can also continuously optimize models based on real-time feedback, better addressing changes in user preferences.

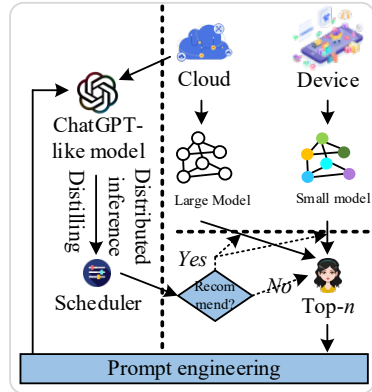

**Fig. 6.** A Research Framework for Device-Cloud Collaborative RSs Based on Fine-Tuning ChatGPT-Like Large Language Models

### 5.3.3. Reinforcement: Utilizing deep reinforcement learning to enhance causal inference

Causal relationships are central to RSs. Understanding and modeling these relationships are crucial for reducing biases, improving interpretability, and impacting recommendation effectiveness and user experience. This study finds that causal inference techniques have significant potential to bridge the gap between correlations and causation in RSs. However, challenges remain in addressing complex causal relationships and dynamic environments, such as poor causal discovery capabilities and a lack of flexibility. Combining deep learning and reinforcement learning to automate causal discovery could be a valuable research direction for enhancing the logic and interpretability of recommendation services. Deep learning excels in pattern recognition and feature extraction, uncovering hidden patterns and regularities in historical interaction data to support automated causal discovery [114]. Meanwhile, reinforcement learning can create state spaces, action spaces, and reward mechanisms, using agent-environment interactions to learn causal rules through trial and error, thus optimizing decision strategies to handle complex causality and uncertainty in the recommendation process.

Using actor-critic reinforcement learning as an example (Fig.7), Transformer can serve as the actor network, handling high-dimensional feature representations of causal relationships in user behavior policies. The critic network evaluates these actions based on the environment state and reward signals, updating recommendation strategies through interactive learning.

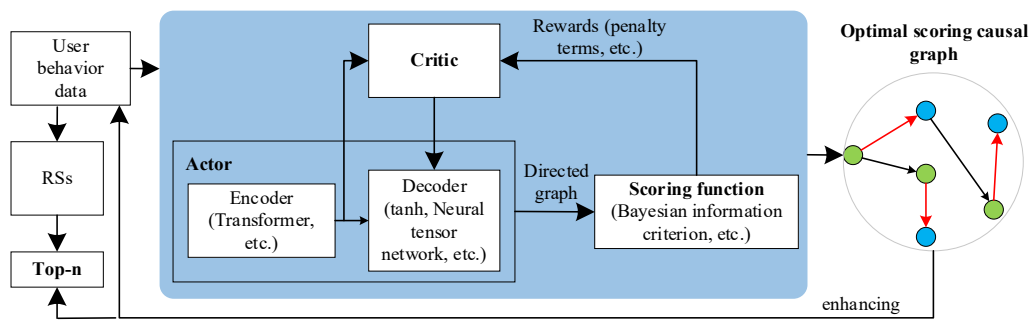

**Fig. 7.** A research Framework for RSs Enhancing Causal Inference Through Actor-Critic Reinforcement Learning

### 5.3.4. Balancing: Training multi-task models based on probability distributions

Multi-task collaboration is essential for RSs to understand user-item relationships comprehensively, reduce recommendation errors, and enhance system robustness. This study finds that multi-task environments involve both information crossover and complementarity, as well as interference and competition among tasks. Current multi-task models and parameter sharing mechanisms fail to balance these factors, optimize collaboration, and reduce recommendation conflicts effectively. Designing a composite function that learns from the overall distribution of relevance judgments could effectively coordinate multiple recommendation tasks. Different ranking models might rate the same items very differently due to selection biases or varying interpretations of objectives. While traditional composite functions can reduce this bias, they often lose critical information. Research in AI indicates that ranking models can improve ranking by using all labels collected during data annotation [115], as item rating is essentially a form of ranking.

For example, a multi-task learning model based on probability distribution loss functions (Fig.8) can probabilistically transform the outputs of different ranking models for various recommendation goals. Using ApproxNDCG loss to design new loss functions based on probability distributions, such as KL divergence, allows the composite model to fully utilize all information from the ranking models.

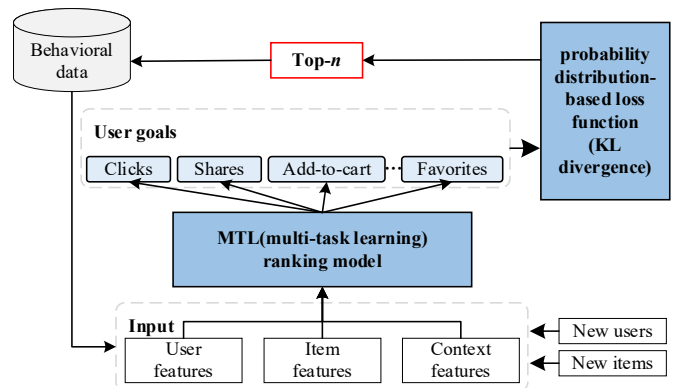

**Fig. 8.** A Research Framework for Multi-Task Learning RSs based on Probability Distribution Loss Function

## 5.4. Evaluation Perspective

### 5.4.1. Customization: Utilizing cross-temporal strategies to partition datasets

Avoiding the use of unavailable future data during offline training is crucial for ensuring that the model accurately reflects historical situations and improving the reliability and usability of RSs. This study finds that random splitting, leave-one-out splitting, and time-point splitting, commonly used in research, suffer from varying degrees of offline data leakage. Although time-aware RSs address this issue to some extent, using the entire training set (global temporal sequences) as a whole inevitably leads to data leakage. Designing data partitioning methods that consider both local and global temporal sequences to prevent offline data leakage could be a valuable research direction. User behaviors are often continuous and sequential; incorporating this into data partitioning can help models better learn behavior patterns and reduce dependence on local future data during training. Additionally, considering global temporal trends can improve the generalization ability of recommendation models in large-scale or emerging scenarios and reduce computational costs.

Using a cross-temporal partitioning method as an example (Fig.9), interaction data can be cross-sorted and test instances sampled from both local and global time dimensions. Models can then be trained in sequential or batch learning [99, 116] to generate recommendations.

### 5.4.2. Implementation: comprehensive evaluation of recommendation objectives throughout the lifecycle

Comprehensive consideration of multiple performance aspects is crucial for understanding RS performance, adapting to changing user needs, and supporting technological development. This study finds that super-accuracy metrics, such as novelty and content safety, are gaining attention. However, their effectiveness and how to balance them with accuracy metrics across different scenarios, tasks, and users need further research. A comprehensive evaluation of RSs throughout their lifecycle could be a valuable research direction for balancing different recommendation objectives. RS performance depends not only on specific points in time but also on factors like data updates, changes in user behavior, and technological evolution. Evaluating RSs at different lifecycle stages can provide a more complete understanding of their performance, aiding decision-making. Additionally, intelligent algorithms and data analysis can dynamically adjust and optimize system adaptability and efficiency in real time. With diverse application scenarios and user needs, a single performance metric is insufficient. Thus, multi-dimensional and multi-perspective evaluation of RSs is important to enhance user experience.

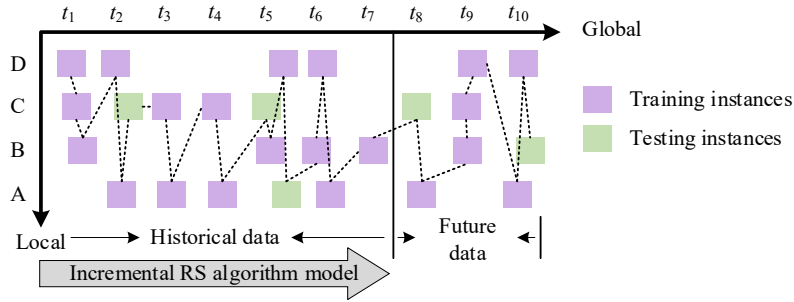

**Fig. 9.** A Training Set and Testing Set Splitting Strategy Across Local and Global Temporal

For example, considering the entire software lifecycle (Fig.10), RS performance can be evaluated based on four contextual dimensions: recommendation scenarios, task types, experiment types, and target groups. This should be done at different RS lifecycle stages, including requirement analysis, system design, coding and implementation, system testing, deployment and delivery, and operation and maintenance [117].

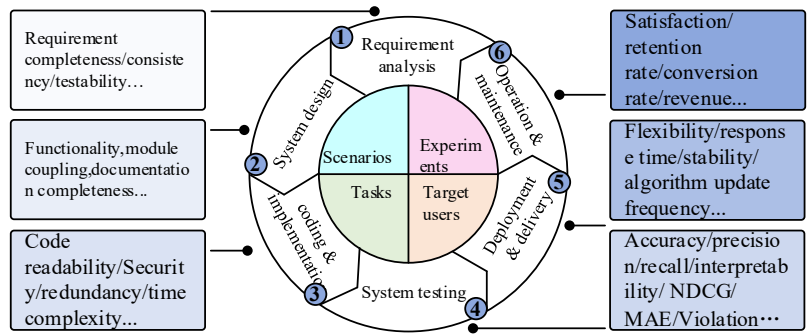

**Fig. 10.** An Evaluation Framework for Multi-Objective RSs Based on the Full Software Lifecycle

## 6. Conclusion

In recent years, RSs have become essential for alleviating user information and cognitive overload, leading to various content-based, knowledge-based, and context-based recommendation technologies and industrial-grade systems. However, rapid technological advancements (such as pre-trained large language models and augmented reality), application environments (like Metaverse and IoT), and diverse user needs present new challenges for RSs. This study systematically reviews the key development challenges in three areas: data, algorithms, and evaluation. It also proposes potential expansions, including integrating physiological signals in multimodal modeling, using user information behavior to defend against data poisoning, evaluating generative recommendations through social experiments, fine-tuning pre-trained large models to manage device-cloud resources, enhancing causal inference with deep reinforcement learning, training multi-task models based on probability distributions, using cross-temporal strategies for dataset partitioning, and comprehensively evaluating recommendation objectives throughout the lifecycle. These insights aim to guide researchers in exploring, studying, and applying RSs in the context of the new liberal arts.

This study has certain limitations. To enhance the feasibility of the systematic review, appropriate and specific constraints were established, including using Web of Science, ScienceDirect, and SpringerLink databases, as well as arXiv and Google Scholar, as the primary sources for paper retrieval, and limiting papers to English. However, from a completeness perspective, collecting papers from as many sources and languages as possible would indeed benefit the systematic review. Therefore, future research could aim to retrieve and collect multi-language RSs studies from additional databases, allowing for a deeper exploration of cutting-edge RS technologies and innovations across different fields, thus expanding the understanding of RS development globally.

**Author Contribution:** All authors contributed equally to the main contributor to this paper. All authors read and approved the final paper.

**Funding:** This research received no external funding.

**Conflicts of Interest:** The authors declare no conflict of interest.